\documentclass{article}
\usepackage{amsmath}
\usepackage{amssymb}

\input{tcilatex}
\begin{document}

\title{The complementary group of proper motions of the Minkowski metric in
an arbitrary dimension space.}
\author{Popov N. N. \\
nnpopov@mail.ru \\
Computer Centre of the Russian Academy of Sciences}
\date{}
\maketitle

\bigskip

\begin{center}

Abstract
\end{center}

\begin{quotation}
It is shown that the Poincare group which is a semidirect product of the
group of translations and the Lorentz group, is not a single physicaly
important group of proper motions of Minkowski metric. The complementary
group of proper motions of the metric in a class of noninertial reference
system has been found.


\bigskip
\end{quotation}

\begin{center}
{\Large Introduction}
\end{center}

The whole group of all proper transformations of Minkowski pseudo-Euclidean
space $R_{1,3}^{4}$, which leaves the metric invariant, in other words the
whole group of proper motions of pseudo-Euclidean space, is a semidirect
product $T^{4}\rtimes SO(1,3)$ of the four-dimensional group of translations
$T^{4}$ and the proper pseudo-orthogonal group $SO(1,3).$ At present we
identify the pseudo-orthogonal group $SO(1,3)$ with the Lorentz group of
proper linear pseudo-orthogonal transformations

\begin{quotation}
\begin{equation}
dx^{0}=dx^{\prime 0}ch\psi +dx^{\prime i}sh\psi ,
\end{equation}
\end{quotation}

\begin{equation*}
dx^{i}=dx^{\prime 0}sh\psi +dx^{\prime i}ch\psi ,
\end{equation*}

where $(x^{\prime 0},x^{\prime 1},x^{\prime 2},x^{\prime
3}),(x^{0},x^{1},x^{2},x^{3})$ are pseudo-Euclidean coordinates of a point
before and after the transformation, $i=1,2,3.$

In terms of special relativity theory the Lorentz transformation provides
the transition from one inertial reference system $(x^{0},x^{1},x^{2},x^{3})$
to another one $(x^{\prime 0},x^{\prime 1},x^{\prime 2},x^{\prime 3})$ which
moves at speed $v$ along axis $x^{i}$, and it retains the differential
quadratic form $dx^{0^{2}}-dx^{1^{2}}-dx^{2^{2}}-dx^{3^{2}}.$ Hence the
Lorentz transformation may be rewritten in the following form:

\begin{equation}
dx^{0}=\dfrac{dx^{\prime 0}+\dfrac{v}{c}dx^{\prime i}}{\sqrt{1-\left( \dfrac{%
v}{c}\right) ^{2}}},
\end{equation}%
\begin{equation*}
dx^{i}=\frac{dx^{\prime i}+\dfrac{v}{c}dx^{\prime 0}}{\sqrt{1-\left( \dfrac{v%
}{c}\right) ^{2}}},
\end{equation*}

where $ch\psi =\dfrac{1}{\sqrt{1-\left( \dfrac{v}{c}\right) ^{2}}},$ $c$ is
a certain constant which special relativity theory interprets as the light
speed.

The semidirect product of the group of translations and the Lorentz group
represents the Poincare group. Hence the Poincare group consists of a group
of proper motions of Minkowski metric in the class of all inertial reference
systems. It would be wrong to say that the Lorentz group coincides with the
group of all proper pseudo-orthogonal transformations $SO(1,3),$ or to
identify the Poincare group with the $T^{4}\rtimes SO(1,3)$ group. In fact,
the Lorentz group is a subgroup of $SO(1,3)$. Moreover, the $SO(1,3)$ group
contains another phisicaly important proper motions group of Minkowski
metric in a class of noninertial frames of reference, which is surprising
enough. Construction of this group provides the subject of this paper.

Taking into account the wide interest in the Minkowski spaces of large
dimension in, for instance, theories of strings, superstrings, supergravity
etc., we consider it advisable to observe the problem for Minkowski spaces $%
R_{1,N}^{N+1}$ with arbitrary dimension of $N\geq 2.$

\bigskip

{\Large 1. Construction of a complementary\ group of proper
pseudo-orthogonal transformations in }${\Large R}_{1,2}^{3}${\Large .}

We will start with the simple case of three-dimensional pseudo-Euclidean
space, in which in addition to Lorentz group, there exists another group of
proper motions of Minkowski metric.

Any proper rotation in a three-dimensional pseudo-Euclidean space $%
R_{1,2}^{3}$, i.e. any proper orthogonal or pseudo-orthogonal transformation
which preserves the coordinates origin, can be decomposed into three
rotations\ in planes $\{x^{1}x^{2}\},$ $\{x^{0}x^{1}\},$ $\{x^{0}x^{2}\},$
and one rotation in the space $R_{1,2}^{3}=\{x^{0}x^{1}x^{2}\}$ itself,
which cannot be reduced to the previous ones. The first rotation modifies
the space coordinates only and corresponds to the space rotations. The
second and third rotations act in pseudo-Euclidean planes and correspond to
the proper pseudo-orthogonal rotations or, which is the same, the Lorentz
transformations of the form (1) or (2). Now we will observe in detail the
extra rotation in the $R_{1,2}^{3}.$

The desired transformation should leave invariant the differential quadratic
form%
\begin{equation}
dx^{0^{2}}-dx^{1^{2}}-dx^{2^{2}},
\end{equation}

or the form equivalent to it, in the polar coordinate system%
\begin{equation}
dx^{0^{2}}-dr^{2}-r^{2}d\varphi ^{2},
\end{equation}

where $x^{1}=r\cos \varphi ,$ $x^{2}=r\sin \varphi .$

Let $(x^{0},r,\varphi )$ be coordinates of a point $s$ in the
three-dimensional pseudo-Euclidean subspace $\{x^{0}x^{1}x^{2}\}.$ We
require that the transformations from the desired group $\mathbf{G}$ leave
unchanged the radial coordinate $r$ of the point $s.$ A subset in $%
\{x^{0}x^{1}x^{2}\}$ which consists of all point of $\mathbf{G}s$ type,
forms an orbit of element $s$ with respect to the group $\mathbf{G}$ and is
denoted $G_{r}(s).$ If $s_{1}$ and $s_{2}$ are two points from $%
\{x^{0}x^{1}x^{2}\}$ which have the same parameter $r,$ then $%
G_{r}(s_{1})=G_{r}(s_{2})$, i.e. orbits of two different points are only
defined by the radial coordinate $r$ of these points. Hence we will use
symbol $G_{r}$ to denote an orbit. If $r_{1}\neq r_{2},$ then $G_{r_{1}}\cap
G_{r_{2}}\subset \oslash ,$ which means that the orbits of two different
points either coincide, or do not intersect. A Pseudo-Euclidean subspace $%
\{x^{0}x^{1}x^{2}\}$ is a union of pairwise not intersecting orbits $%
G_{r}:\{x^{0}x^{1}x^{2}\}=\cup _{0<r<\infty }G_{r}.$

Let $f_{r}$ be the homomorphism of a group $\mathbf{G}$ into group $\mathbf{G%
}_{r}$ effective in orbit $G_{r}:f_{r}(\mathbf{G})=\mathbf{G}_{r}.$
According to the definition of effectiveness of a group $\mathbf{G}$ in an
orbit $G_{r},$ we have $\mathbf{G}s=\mathbf{G}_{r}s,$ when $s\in G_{r}.$

Now we directly construct the group $\mathbf{G}_{r}.$ Let the differentials
of coordinates $x^{0},\varphi $ of a point $s(x^{0},r,\varphi )$ be subject
to linear transformation $a_{\omega }:(dx^{0},d\varphi )\rightarrow
(dx^{\prime 0},d\varphi ^{\prime })$ of the form:

\begin{equation}
dx^{0}=\frac{dx^{\prime 0}+\dfrac{r^{2}\omega }{c}d\varphi ^{\prime }}{\sqrt{%
1-(\dfrac{r\omega }{c})^{2}}},
\end{equation}

\begin{equation*}
d\varphi =\frac{d\varphi ^{\prime }+\dfrac{\omega }{c}dx^{\prime 0}}{\sqrt{%
1-\left( \dfrac{r\omega }{c}\right) ^{2}}},
\end{equation*}

where $\omega $ is the angular velocity of a circle of radius $r$ in a plane
$\{x^{1}x^{2}\}$ relative to the origin of coordinates, $|\omega |<\dfrac{c}{%
r}.$

It is easy to see that the linear transformation (5) leaves invariant the
quadratic form (4) and is the element of the group $\mathbf{G}_{r}.$
Obviously, the element $a_{0}$ corresponds to unit I of the group $\mathbf{G}%
_{r}.$ The element $a_{-\omega }$ is identified as an element inverse to $%
a_{\omega }\in \mathbf{G}_{r},$ i.e. $(a_{\omega })^{-1}=a_{-\omega }.$ If $%
a_{\omega _{1}},a_{\omega _{2}}\in \mathbf{G}_{r},$ then their group product
may be determined as

\begin{equation*}
a_{\omega _{1}}\cdot a_{\omega _{2}}=a_{\frac{\omega _{1}+\omega _{2}}{%
1+\left( \frac{r}{c}\right) ^{2}\omega _{1}\omega _{2}}}.
\end{equation*}

If $|\omega _{1}|,|\omega _{2}|<\dfrac{c}{r},$ it immediately follows from
the group multiplication that $|\dfrac{\omega _{1}+\omega _{2}}{1+\left(
\frac{r}{c}\right) ^{2}\omega _{1}\omega _{2}}|<\dfrac{c}{r}.$

The group $\mathbf{G}_{r}$ can formally be supplemented by adding two
elements $a_{-\frac{c}{r}}$ and $a_{\frac{c}{r}}.$

The replenished group will be stationary relative to the associated elements 
$a_{-\frac{c}{r}}$ and $a_{\frac{c}{r}}.$ Indeed, for any $a_{\omega },$ $%
|\omega |\leq \dfrac{c}{r}$ we have $a_{\omega }\cdot a_{\pm \frac{c}{r}%
}=a_{\pm \frac{c}{r}}.$

In terms of special relativity theory it means that the linear velocity of a
circular motion cannot exceed the light speed $c$ and that the light speed
is identical in all steadily rotating\ noninertial reference systems.

If we introduce the variable $y=r\varphi ,$ the transformation (5) may be
rewritten in the following form:

\begin{equation}
dx^{0}=\frac{dx^{\prime 0}+\frac{v}{c}dy^{\prime }}{\sqrt{1-(\frac{v}{c})^{2}%
}},\text{ }dy=\frac{dy^{\prime }+\frac{v}{c}dx^{\prime 0}}{\sqrt{1-(\frac{v}{%
c})^{2}}},
\end{equation}

where $v=r\omega .$

If we compare the Lorentz transformations (2) with the pseudo-orthogonal
transformations (6), it is not hard to observe their formal structural
similarity, written in different reference systems. However, there is much
difference between them. The Lorentz transform describes the transition
between inertial reference systems which move uniformly and straight
relative to each other, yet the transformation (6) describes the transition
between noninertial systems rotating uniformly in circles with different
angular velocities. This difference becomes clear when we go over from the
polar coordinate system to the pseudo-Euclidean one, a considerably
nonlinear character of the transformations emerges. Indeed, if we go over to
the pseudo-Euclidean coordinate system $x^{0},x^{1},x^{2},$ then the
transformation (6) assumes\ the form

\begin{equation*}
dx^{0}=\frac{1}{\sqrt{1-(\frac{v}{c})^{2}}}dx^{\prime 0}-\frac{\omega }{c}%
\frac{x^{\prime 2}}{\sqrt{1-(\frac{v}{c})^{2}}}dx^{\prime 1}+\frac{\omega }{c%
}\frac{x^{\prime 1}}{\sqrt{1-(\frac{v}{c})^{2}}}dx^{\prime 2},
\end{equation*}

\begin{equation}
dx^{1}=-\frac{v}{c}\frac{\sin \varphi }{\sqrt{1-(\frac{v}{c})^{2}}}%
dx^{\prime 0}+\frac{\omega }{v}Adx^{\prime 1}+\frac{\omega }{v}Bdx^{\prime
2},
\end{equation}%
\begin{equation*}
dx^{2}=\frac{v}{c}\frac{\cos \varphi }{\sqrt{1-(\frac{v}{c})^{2}}}dx^{\prime
0}+\frac{\omega }{v}Cdx^{\prime 1}+\frac{\omega }{v}Ddx^{\prime 2},
\end{equation*}

where $\varphi =\dfrac{\arccos \frac{x^{\prime 1}}{\sqrt{x^{\prime
1^{2}}+x^{\prime 2^{2}}}}+\frac{\omega }{c}x^{\prime 0}}{\sqrt{1-\left( 
\frac{v}{c}\right) ^{2}}},$ $A=x^{\prime 1}\cos \varphi +\frac{x^{\prime
2}\sin \varphi }{\sqrt{1-\left( \frac{v}{c}\right) ^{2}}},$ $B=x^{\prime
2}\cos \varphi -\frac{x^{\prime 1}\sin \varphi }{\sqrt{1-\left( \frac{v}{c}%
\right) ^{2}}},$ $C=x^{\prime 1}\sin \varphi -\frac{x^{\prime 2}\cos \varphi 
}{\sqrt{1-\left( \frac{v}{c}\right) ^{2}}},$ $D=x^{\prime 2}\sin \varphi +%
\frac{x^{\prime 1}\cos \varphi }{\sqrt{1-\left( \frac{v}{c}\right) ^{2}}}.$

The whole group of all proper pseudo-orthogonal transformations $SO(1,2)$ in
a space $R_{1,2}^{3}$ is thus generated by elements from the Lorentz group
(2) and by transformations of form (7) from the group $\mathbf{G}_{r}$ in
the pseudo-Euclidean coordinate system.

\bigskip {\Large 2. The complementary group of proper pseudo-orthogonal
transformations in }${\Large R}_{1,2n}^{2n+1}$

In the previous paragraph we have reviewed a simple case of odd-dimensional
Minkowski space $R_{1,2n}^{2n+1}$ with $n=1,$ in which the construction of
additional motion group was favoured by a lucky choice of coordinate system.
The analogous construction may also be done in the general case.

Any element $A$ of the maximum subgroup $SO(2n),$ constituent of $SO(1,2n)$
group, according to the known result in linear algebra [1] is represented as
a block-diagonal matrix%
\begin{equation}
A=\left( 
\begin{array}{ccccc}
\square & 0 & 0 & 0 & 0 \\ 
0 & \ddots & 0 & 0 & 0 \\ 
0 & 0 & \square & 0 & 0 \\ 
0 & 0 & 0 & \ddots & 0 \\ 
0 & 0 & 0 & 0 & \square%
\end{array}%
\right) ,
\end{equation}

where the k$^{th}$diagonal block has the form $\left( 
\begin{array}{cc}
\cos \varphi _{k} & \sin \varphi _{k} \\ 
-\sin \varphi _{k} & \cos \varphi _{k}%
\end{array}%
\right) .$

On the basis of (8) we will introduce a biharmonic coordinate system [2] $%
r,\varphi _{1},\ldots ,\varphi _{n},\theta _{1},\ldots ,\theta _{n-1},$
which divides all Cartesian coordinates $x^{1},\ldots ,x^{2n}$ from
Euclidean subspace $R^{2n}$ into pairs $\left( z^{k},y^{k}\right) ,$ where $%
z^{k}=x^{2k-1},$ $y^{k}=x^{2k},$ $k=1,\ldots ,n,$ the $x^{0}$ coordinate
remains unchanged.

Suppose that%
\begin{eqnarray}
z^{1} &=&r\cos \varphi _{1}\sin \theta _{1}\ldots \sin \theta _{n-2}\sin
\theta _{n-1},  \notag \\
y^{1} &=&r\sin \varphi _{1}\sin \theta _{1}\ldots \sin \theta _{n-2}\sin
\theta _{n-1},  \notag \\
&&.............................  \notag \\
z^{k} &=&r\cos \varphi _{k}\sin \theta _{1}\ldots \sin \theta _{n-k}\cos
\theta _{n-k+1}, \\
y^{k} &=&r\sin \varphi _{k}\sin \theta _{1}\ldots \sin \theta _{n-k}\cos
\theta _{n-k+1},  \notag \\
&&.............................  \notag \\
z^{n} &=&r\cos \varphi _{n}\cos \theta _{1},  \notag \\
y^{n} &=&r\sin \varphi _{n}\cos \theta _{1},  \notag
\end{eqnarray}

where $k=2,\ldots ,n.$

The differential quadratic form%
\begin{equation*}
dx^{0^{2}}-dx^{1^{2}}-\ldots
-dx^{2n^{2}}=dx^{0^{2}}-dz^{1^{2}}-dy^{1^{2}}-\ldots -dz^{n^{2}}-dy^{n^{2}},
\end{equation*}

which remains invariant by the Lorentz group of transformations, taking the
following form in biharmonic coordinate system%
\begin{equation}
dx^{0^{2}}-dr^{2}-r_{1}^{2}d\varphi _{1}^{2}-\ldots -r_{n}^{2}d\varphi
_{n}^{2}-r^{2}\left( d\theta _{1}^{2}+\sin ^{2}\theta _{1}\left( d\theta
_{2}^{2}+\sin ^{2}\theta _{2}\left( \ldots +\sin ^{2}\theta _{n-2}d\theta
_{n-1}^{2}\right) \right) \ldots \right) ,
\end{equation}

where $r_{1}=r\sin \theta _{1}\ldots \sin \theta _{n-1},\ldots ,$ $%
r_{k}=r\sin \theta _{1}\ldots \sin \theta _{n-k}\cos \theta _{n-k+1},$ $%
k=2,\ldots ,n,$ i.e. the following condition is fulfilled 
\begin{equation}
r_{1}^{2}+\ldots +r_{n}^{2}=r^{2}.
\end{equation}

Let $r_{1},\ldots ,r_{n}$ be fixed, then the differential quadratic form
(10) remains invariant relative to $n$ one-parameter subgroups of proper
pseudo-orthogonal transformations, where the k$^{th}$ subgroup consists of
the following transformations%
\begin{eqnarray}
dx^{0} &=&\frac{dx^{\prime 0}+\frac{r_{k}^{2}\omega }{c}d\varphi
_{k}^{\prime }}{\sqrt{1-\left( \frac{r_{k}\omega }{c}\right) ^{2}}}, \\
d\varphi _{k} &=&\frac{d\varphi _{k}^{\prime }+\frac{\omega }{c}dx^{\prime 0}%
}{\sqrt{1-\left( \frac{r_{k}\omega }{c}\right) ^{2}}}.  \notag
\end{eqnarray}

Obviously, the required invariability for parameters $r_{1},\ldots ,r_{n}$
by pseudo-orthogonal transformations (12) is equivalent to the requirement
of constancy for $\theta _{1},\ldots ,\theta _{n-1},$ which provides the
invariance for the differential form 
\begin{equation*}
r^{2}\left( d\theta _{1}^{2}+\sin ^{2}\theta _{1}\left( \ldots +\sin
^{2}\theta _{n-2}d\theta _{n-1}^{2}\right) \ldots \right) .
\end{equation*}
The invariance of the quadratic form $dx^{0^{2}}-r_{1}^{2}d\varphi
_{1}^{2}-\ldots -r_{n}^{2}d\varphi _{n}^{2}$ is due to the very type of
transformations (12). Hence the group of all proper pseudo-orthogonal
transformations $G_{r},$ which leaves the radial parameter $r$ unchanged, is
generated by all possible pseudo-orthogonal transformations of form (12)
from $n$ subgroups $G_{r_{1}},\ldots ,$ $G_{r_{n}}$ which correspond to
fixed arbitrary sets $r_{1},\ldots ,r_{n},$ satisfying the condition (11).

The expansional group of proper pseudo-orthogonal transformations in $%
R_{1,2n}^{2n+1}$ is thus generated by the Lorentz group and group $G_{r}.$

\bigskip {\Large 3. The complementary group of the proper pseudo-orthogonal
transformations in }${\Large R}_{1,2n+1}^{2(n+1)}$

In the case of even-dimensional Minkowski space, any element $A$ of the
maximum orthogonal subgroup $SO(2n+1)$ from group $SO(1,2n+1)$ may be
represented in a block-diagonal form%
\begin{equation}
A=\left( 
\begin{array}{cccc}
\square & 0 & 0 & 0 \\ 
0 & \ldots & 0 & 0 \\ 
0 & 0 & \square & 0 \\ 
0 & 0 & 0 & 1%
\end{array}%
\right) ,
\end{equation}

where the k$^{th}$ diagonal block has the same form as in the relation (8).

Dividing the Cartesian coordinates $x^{1},\ldots ,x^{2n}$, as in the
previous case, into pairs $(z^{k},y^{k}),$ where $x^{2k-1}=z^{k},$ $%
x^{2k}=y^{k},$ $k=1,\ldots ,n,$ and assuming $x^{2n+1}=z^{n+1},$ we will
introduce the biharmonic coordinate system $r,\varphi _{1},\ldots ,\varphi
_{n},\theta _{1},\ldots ,\theta _{n}$ so that%
\begin{eqnarray}
z^{1} &=&r\cos \varphi _{1}\sin \theta _{1}\ldots \sin \theta _{n},  \notag
\\
y^{1} &=&r\sin \varphi _{1}\sin \theta _{1}\ldots \sin \theta _{n}, \\
&&....................  \notag \\
z^{k} &=&r\cos \varphi _{k}\sin \theta _{1}\ldots \sin \theta _{n-k+1}\cos
\theta _{n-k+2},  \notag \\
y^{k} &=&r\sin \varphi _{k}\sin \theta _{1}\ldots \sin \theta _{n-k+1}\cos
\theta _{n-k+2},  \notag \\
&&......................  \notag \\
z^{n} &=&r\cos \varphi _{n}\sin \theta _{1}\cos \theta _{2},  \notag \\
y^{n} &=&r\sin \varphi _{n}\sin \theta _{1}\cos \theta _{2},  \notag \\
z^{n+1} &=&r\cos \theta _{1}.  \notag
\end{eqnarray}

The differential quadratic form

$dx^{0^{2}}-dx^{1^{2}}-\ldots
-dx^{2n+1^{2}}=dx^{0^{2}}-dz^{1^{2}}-dy^{1^{2}}-\ldots
-dz^{n^{2}}-dy^{n^{2}}-dz^{n+1^{2}}$ is invariant relative to the Lorentz
group of transformations and in the biharmonic coordinate system (14) it
assumes the form

\begin{equation*}
dx^{0^{2}}-dr^{2}-r_{1}^{2}d\varphi _{1}^{2}-\ldots -r_{n}d\varphi
_{n}^{2}-r^{2}\left( d\theta _{1}^{2}+\sin ^{2}\theta _{1}\left( d\theta
_{2}^{2}+\sin ^{2}\theta _{2}\left( \ldots +\sin ^{2}\theta _{n-1}d\theta
_{n}^{2}\right) \right) \ldots \right) ,
\end{equation*}

where $r_{1}=r\sin \theta _{1}\ldots \sin \theta _{n},\ldots ,r_{k}=r\sin
\theta _{1}\sin \theta _{n-k+1}\cos \theta _{n-k+2},$ $k=2,\ldots ,n,$ i.e.
the following condition is fulfilled 
\begin{equation}
r_{1}^{2}+\ldots +r_{n}^{2}=r^{2}\sin ^{2}\theta _{1}.
\end{equation}

The form remains invariant relative to any pseudo-orthogonal transformations
of the form (12) and is identical with the form (10) with the only
difference that parameters $r_{1},\ldots ,r_{n}$ satisfy the condition (15)
and not (12) as in the case of the odd-dimensional spaces.

These results be summed up as the following statement:

\textbf{Theorem.} \textit{The subgroup of proper pseudo-orthogonal
transformations from \ }$SO(1,N)$\textit{\ of Minkowski space }$%
R_{1,N}^{N+1} $\textit{\ is generated by various transformations from the
Lorentz group of the form}%
\begin{equation*}
dx^{0}=\frac{dx^{\prime 0}+\frac{v}{c}dx^{\prime i}}{\sqrt{1-(\frac{v}{c}%
)^{2}}},\text{ }dx^{i}=\frac{dx^{\prime i}+\frac{v}{c}dx^{\prime 0}}{\sqrt{%
1-(\frac{v}{c})^{2}}},\text{ }i=1,\ldots ,N,
\end{equation*}%
\textit{relative to the pseudo-Euclidean coordinate system }$x^{0},\ldots
,x^{N},$\textit{\ and by various transformations relative to the biharmonic
coordinate system }$x^{0},r,\varphi _{1},\ldots ,\varphi _{\lbrack \frac{N}{2%
}]},\theta _{1},\ldots ,\theta _{\lbrack \frac{N}{2}]}$\textit{\ from group }%
$G_{r}$\textit{\ of the form}%
\begin{equation*}
dx^{0}=\frac{dx^{\prime 0}+\frac{r_{k}^{2}\omega }{c}d\varphi _{k}^{\prime }%
}{\sqrt{1-(\frac{r_{k}\omega }{c})^{2}}},\text{ }d\varphi _{k}=\frac{%
d\varphi _{k}^{\prime }+\frac{\omega }{c}dx^{\prime 0}}{\sqrt{1-(\frac{%
r_{k}\omega }{c})^{2}}},\text{ }k=1,\ldots ,[\frac{N}{2}],
\end{equation*}%
\textit{with fixed }$r_{1},\ldots ,r_{[\frac{N}{2}]},r,$\textit{\ satisfying
the following condition }$r_{1}^{2}+\ldots +r_{[\frac{N}{2}]}^{2}=r^{2}$%
\textit{\ if N is even and }$r_{1}^{2}+\ldots +r_{[\frac{N}{2}%
]}^{2}=r^{2}\sin ^{2}\theta _{1}$\textit{\ if N is odd.}

{\Large 4. The Maxwell equations invariance relative to the motion group }$%
{\Large G}_{{\Large r}}$

We will show that the Maxwell equations%
\begin{gather}
\frac{\partial F_{ij}}{\partial x^{k}}+\frac{\partial F_{ki}}{\partial x^{j}}%
+\frac{\partial F_{jk}}{\partial x^{i}}=0, \\
\frac{\partial F^{ij}}{\partial x^{j}}=J^{i},\text{ \ \ }i,j=0,1,...,3
\end{gather}

appear invariant relative to the complementary transformation group $G_{r}$
in the four-dimentional Minkowski space $R_{1,3}^{4},$ where 
\begin{eqnarray}
F^{ij} &=&F_{ij}\text{ for }i,j=1,2,3 \\
F^{oj} &=&-F_{oj}\text{ for }j=1,2,3.  \notag
\end{eqnarray}

That is, we want to show that an arbitrary transformation of coordinates $%
x^{0},x^{1},x^{2},x^{3}$ from the group $G_{r}$ of kind (7) in new
coordinates $x^{\prime 0},x^{\prime 1},x^{\prime 2},x^{\prime 3},$ leaves
invariant the Maxwell equations (16), (17) relative to transformed tensors $%
F_{i^{\prime }j^{\prime }}=\frac{\partial x^{i}}{\partial x^{i^{\prime }}}%
\frac{\partial x^{j}}{\partial x^{j^{\prime }}}F_{ij},$ $F^{i^{\prime
}j^{\prime }}=\frac{\partial x^{i^{\prime }}}{\partial x^{i}}\frac{\partial
x^{j^{\prime }}}{\partial x^{j}}F^{ij},$ and that the condition (18) is
fulfilled.

In fact, having in mind that the Minkowski metric remains invariant during
the pseudo-orthogonal transformations of kind (7), the correlation (18)
remain unchanged for the transformed tensors $F^{i^{\prime }j^{\prime }}$
and $F_{i^{\prime }j^{\prime }}.$

Equation (16) remains invariant after any continuously differentiable
nondegenerate transformation of coordinates, which is following from the
structure of this equation, which represents the Bianchi identity. Further,
counting that $\frac{\partial }{\partial x^{j}}$ is transformed as a vector
during any pseudo-orthogonal coordinate transformations, of kind (7) in
particular, we are having $\frac{\partial }{\partial x^{j}}=\frac{\partial
x^{m^{\prime }}}{\partial x^{j}}\frac{\partial }{\partial x^{m^{\prime }}}.$
Then the left side of the Eq (17) can be represented in the following way:%
\begin{equation*}
\frac{\partial F^{ij}}{\partial x^{j}}=\frac{\partial x^{m^{\prime }}}{%
\partial x^{j}}\frac{\partial }{\partial x^{m^{\prime }}}\frac{\partial x^{i}%
}{\partial x^{k^{\prime }}}\frac{\partial x^{j}}{\partial x^{l^{\prime }}}%
F^{k^{\prime }l^{\prime }}=\frac{\partial x^{i}}{\partial x^{k^{\prime }}}%
\frac{\partial F^{k^{\prime }l^{\prime }}}{\partial x^{l^{\prime }}}
\end{equation*}

and the equation (17) assumes the following form in the new coordinates:%
\begin{equation*}
\frac{\partial F^{k^{\prime }l^{\prime }}}{\partial x^{l^{\prime }}}%
=J^{k^{\prime }},\text{ where }J^{k^{\prime }}=\frac{\partial x^{k^{\prime }}%
}{\partial x^{i}}J^{i}.
\end{equation*}

Thus, we consider the Eq (17) invariance to be proven during the
pseudo-orthogonal thanformations of kind (7).

Using the previos constraction we come to a general conclusion that the
Maxwell equations are invariant relative to all proper motions of Minkowski
metric. Strictly speaking, this conclusion immediately follows from the
proof of Maxwell equations invariance relative to the group $G_{r},$
generalized to the whole group of metric proper motions.

Various combinations of Lorentz transformations, and transformations from
the group $G_{r}$ lead to a rather wide set of possible motions of
non-inertial reference systems, relative to which Maxwell equations retain
their form. Such motions are represented by the uniform cycloidal motion, or
a motion in a helical line along a coordinate axis, or better a uniform
spiral motion on a torus. Thus, every proper transformation from the group $%
SO(1,3)$ has a corresponding continuous motion of a coordinate system,
relative to which Maxwell equations retain their form. For example,
relativistic motion in the spiral line at constant speed $V$ along $X^{3}$
axis and at constant speed $v$ in plane $\{x^{1},x^{2}\}$ in a circle of
radius $r,$ is corresponding to the pseudo-orthogonal coordinate
transformation from the proper motions group of Minkowski metric%
\begin{equation*}
\left( 
\begin{array}{c}
dx^{0} \\ 
dr \\ 
d\varphi \\ 
dx^{3}%
\end{array}%
\right) =\left( 
\begin{array}{cccc}
\xi \eta & 0 & \dfrac{vr}{c}\xi \eta & \dfrac{V}{c}\xi \\ 
0 & 1 & 0 & 0 \\ 
\dfrac{v}{cr}\eta & 0 & \eta & 0 \\ 
\dfrac{V}{c}\xi \eta & 0 & \dfrac{Vvr}{c^{2}}\xi \eta & \xi%
\end{array}%
\right) \left( 
\begin{array}{c}
dx^{\prime 0} \\ 
dr^{\prime } \\ 
d\varphi ^{\prime } \\ 
dx^{\prime 3}%
\end{array}%
\right)
\end{equation*}

where $\xi =\frac{1}{\sqrt{1-\left( \dfrac{V}{c}\right) ^{2}}},$ $\eta =%
\frac{1}{\sqrt{1-\left( \dfrac{v}{c}\right) ^{2}}}.$

This explains why the Maxwell equations are invariant relative to this
thanformation.

\begin{center}
{\Large 5. Discussion and conclusions}
\end{center}

The existence of the extra group of proper motions of the Minkowski metric
in the class of noninertial reference frames gives us a new fundamental
group of Minkowski space symmetry. It occures a question if this fact can
lead us, in the case of four-dimensional Minkowski space, to a revision of
the fundamentals of special relativity theory. Should we require invariance
for all physical theories in respect to the new symmetry group? The fact
that the light speed and the Maxwell electrodynamic equations, as it was
shown above, result invariant relative to various uniformly rotating
non-inertial reference systems makes us feel that such requirement may be
justified. On the other hand, the fact that we can distinguish, among
reference systems uniformly rotating around a common center, a reference
system in a state of complete rest relative to this center, contradicts the
very spirit of special relativity theory.

\bigskip

\begin{center}
\ \ \ \ \ References
\end{center}

\end{document}